\documentclass[conference]{IEEEtran}
\usepackage[utf8]{inputenc}
\usepackage{amsmath,amssymb,amsfonts}
\usepackage{calc}
\usepackage{verbatim,leqno}
\usepackage[normalem]{ulem}
\usepackage{latexsym}
\usepackage{float,epsfig,color}
\usepackage{algorithm}
\usepackage{algorithmicx}
\usepackage{algcompatible}
\usepackage{algpseudocode}
\usepackage{subcaption}
\usepackage[bookmarks,breaklinks]{hyperref}
\usepackage{mdframed}
\usepackage{amsmath,amssymb,amsfonts}
\usepackage{calc}
\usepackage{verbatim,leqno}
\usepackage{latexsym}
\usepackage{caption}
\usepackage{float,epsfig,color}
\usepackage{blindtext}
\usepackage{graphicx}
\usepackage{bbm}
\usepackage{algorithm}
\usepackage{algcompatible}
\usepackage{multirow}
\usepackage{rotating}
\usepackage[nocomma]{optidef}
\usepackage{algpseudocode}
\usepackage{blindtext}
\usepackage{algpseudocode}
{}
{}
\usepackage{blindtext}
\begin{document}

\title{Thermal-aware Workload Distribution for Data Centers with Demand Variations\\
\thanks{Identify applicable funding agency here. If none, delete this.}
}
\author{\IEEEauthorblockN{Somayye Rostami, Douglas G.\ Down, and George Karakostas\\}

\IEEEauthorblockA{Department of Computing and Software \\
McMaster University\\
Hamilton, ON, Canada \\Email:
  \texttt{\{rostas1,downd,karakos\}@mcmaster.ca}\\[1ex]}
}

\maketitle

\begin{abstract}
 Thermal-aware workload distribution is a common approach in the literature for power consumption optimization in data centers. However, data centers also have other operational costs such as the cost of equipment maintenance and replacement. It has been shown that server reliability depends on frequency of their temperature variations, arising from workload transitions due to dynamic demands. In this work, we formulate a nonlinear optimization problem that considers the cost of workload transitions in addition to IT and cooling power consumption. To approximate the solution, we first linearize the problem; the result is a mixed integer programming problem. A modified heuristic is then proposed to approximate the solution of the linear problem. Finally, a Model Predictive Control (MPC) approach is integrated with the proposed heuristics for automatic workload reconfiguration when future demand is not known exactly, but predictions are available. Numerical results show that the proposed schemes are attractive in different settings.
\end{abstract}


\section{Introduction}
\label{int}

\begin{figure*}[!t] 
\begin{equation}\label{p1}
		\begin{aligned}
		 \min \: \: \quad \sum_{k=1}^{K}  F(&v^{(k)}) + \sum_{k=1}^{K} w_k \sum_{i=1}^{n} |\rho_{i}^{(k)}-\rho_{i}^{(k-1)}| +  \sum_{k=1}^{K} \sum_{i=1}^{n} P(\rho_i^{(k)}, t^{(k)}_i) &\\
			  \text{s.t.} \quad		 
			\sum_{i=1}^{n} \rho_{i}^{(k)} & \geq  D_k  \,\qquad\qquad\qquad\qquad\qquad\qquad\qquad\forall k=1,...,K \\			
M(v^{(k)},\rho^{(k)}) & \leq T_{idle}1_{n \times 1}-(T_{idle}-T_{busy})\rho^{(k)}  \: \qquad  \forall k=1,...,K \\			
			v^{(k)} & \geq  V_{LB}   \,\,\,\quad\qquad\qquad\qquad\qquad\qquad\qquad \forall k=1,...,K \\		
			v^{(k)} & \leq  V_{UB} \,\, \,\quad\qquad\qquad\qquad\qquad\qquad\qquad \forall k=1,...,K \\
												 \rho_{i}^{(k)} & \in  \{0,1\} \quad\qquad\qquad\qquad\qquad\qquad\qquad \forall k=1,...,K \quad \forall i=1,...,n 		\end{aligned}
	\end{equation} 
 \hrulefill
\vspace*{4pt}
 \end{figure*}

  Energy consumption of data centers is increasing rapidly, due to growth in demand for internet services and cloud computing tasks. IT and cooling equipment are the main power consumers in a data center \cite{s0}\cite{s1}. Due to heat recirculation effects, thermal-aware workload distribution is necessary to minimize the power consumption while respecting operational temperature constraints \cite{H1}.

  Data centers also have operational costs such as the cost of  equipment maintenance and replacement. The reliability of servers depends on several factors such as their inlet temperature and frequency of temperature variations. In thermal-aware workload distribution, the inlet temperatures are typically bounded by a red-line temperature \cite{M1}. However, when the workload distribution is changed due to dynamic demands, the cost of varying the workload on a server (which we will call switching costs) has not been addressed in the literature. The varying workload leads to temperature variations that can impact the reliability of servers \cite{tem}. To be more precise, most of the approaches solve the thermal-aware workload distribution problem for a fixed demand. Considering time varying demand is beneficial for reducing the switching costs \cite{sw}. In this case, it would be desirable to incorporate demand predictions into the problem. So, we are interested in a thermal-aware workload distribution problem where demand is time varying and the effects of varying the utilizations of servers are taken into account when (re)distributing workload.
  
  While dynamic workload allocation may have costs associated with decreased server lifetimes, workload migration may also be required. This mainly affects the quality of service by imposing a delay on processing when virtual machines migrate between different physical machines \cite{VM1}\cite{VM2}. In this work, we do not address the virtual machine migration problem directly, however, the problem we plan to solve can also consider the migration cost indirectly by adding a penalty for workload transitions. This penalty can lead to decrease the number of migrations.

   There are a few works that consider switching costs in  the workload distribution policy. Most of the literature addresses thermal-aware workload distribution for a constant demand (steady- state) \cite{tac}\cite{pak}\cite{KKT}\cite{jap}\cite{He1}\cite{He2}. In \cite{sw}, switching costs are considered but cooling power consumption is not considered. In \cite{tase}, switching costs are considered in the thermal-aware workload distribution policy, where a transient thermal model is used. A particle-based optimization algorithm is then used to solve the problem. In this work, we formulate a thermal-aware workload distribution problem in discrete time that considers switching costs in addition to IT and cooling power consumption. The proposed problem is a generalized form of the problem introduced in \cite{M3}\cite{our}.

   The problem proposed in \cite{our} is a general power optimization problem with nonlinear cooling power consumption and steady-state thermal model. As a power reduction scenario, they also consider two different red-line temperatures corresponding to idle and fully-utilized servers, respectively. However, the demand value is fixed. The approach proposed to solve nonlinear power optimization problems is to linearize the problem. Depending on the type of variables (continuous or integral) the resulting problem is a linear programming  problem or an (mixed) integer linear programming problem. This approach is especially beneficial because the nonlinear problem is general enough to represent a range of similar problems in this area. The approach is also time efficient because the thermal models may be computationally expensive when they are used to calculate the temperatures, for example if they are based on solving differential equations. Using linear regression  to construct a linear model helps reduce the time complexity. Finally, a heuristic is proposed to approximate the solution of the linear problem that is then applied to the original problem. 
   
   In this work, we also linearize the problem and generalize the heuristic to approximate the solution of the resulting mixed integer programming problem. The solution is then used for the original problem. When demand predictions are available, we integrate a Model Predictive Control (MPC) approach with the proposed heuristic. Using MPC is common in the literature in the presence of transient thermal models \cite{tase}\cite{M3}\cite{MPC1}\cite{MPC2}\cite{MPC3}. In this work, we show that an MPC approach is useful for reducing the size of the problem and for incorporating updates to the predicted demand.
   Our contributions can be listed as follows:

   \begin{itemize}
       \item Generalization of the constant demand problem to a discrete-time, time-varying problem which also considers switching costs
       \item  Generalization of the heuristic proposed in \cite{our} for the resulting mixed integer linear        programming problem
and proving its applicability for the proposed problem
       
       \item Integration of an MPC approach with demand predictions for the proposed heuristic 
       \item Evaluation of the proposed schemes that suggest the potential for significant cost reductions, e.g. when compared to separating the problem into independent instances at each time step
   \end{itemize}

In the remainder of the paper, we first describe the system model and introduce the optimization problem, in Section \ref{sys}. We also linearize the problem. An approximation algorithm to solve the problem is proposed in Section \ref{app}. An MPC approach is introduced in Section \ref{mpc}. Section \ref{eva} covers the evaluation of the proposed schemes for the introduced integer linear programming problem. Concluding remarks are provided in Section \ref{con}.
   

  \begin{figure*}[!t]



			

\begin{equation}\label{pr1}
		\begin{aligned}
		 {\min} \quad  \sum_{j=1}^{m}  v_j &   \\
			  \text{s.t.} \quad			 
			\sum_{i=1}^{n} \rho_i & \geq D \\
						-\sum_{j=1}^{m}A_{l,j} v_{j}{+} 
      	\sum_{i=1}^{n} B_{l,i} \rho_i 	+ a \rho_l 		& \leq  b- E_l   \quad \forall l=1,...,n  \\			 
			v_{j} & \geq  V_{LB}^{(j)}  \: \: \: \quad  \forall j=1,...,m  \\
			v_{j}  & \leq  V_{UB}^{(j)}  \: \: \: \quad  \forall j=1,...,m  \\
												 \rho_i & \in  \{0,1\} \:  \quad   \forall i=1,...,n \\
		\end{aligned}
	\end{equation}


			


			


 \hrulefill
\vspace*{4pt}
\end{figure*}
\begin{figure*}
\begin{equation}\label{p3}
		\begin{aligned}
		\min  \quad \sum_{k=1}^{K}\sum_{j=1}^{m} & v^{(k)}_j +   \sum_{k=1}^{K} w_k  \sum_{i=1}^{n} s_{k,i} +
  (c+d) \sum_{k=1}^{K} \sum_{i=1}^{n} \rho_{i}^{(k)}\\
			\text{s.t.} \quad			 
			\sum_{i=1}^{n} \rho_
   {i}^{(k)} & \geq  D_k \qquad \qquad  \forall k=1,...,K\\			
						-\sum_{j=1}^{m}A_{l,j} v_{k,j}+ \sum_{i=1}^{n}B_{l,i} \rho_
   {i}^{(k)}	+ a \rho_
   {l}^{(k)}		 & \leq b- E_l  \: \, \, \qquad  \forall k=1,...,K \quad \forall l=1,...,n \\			
			 s_{k,i}-\rho_
   {i}^{(k)}+\rho_
   {i}^{(k-1)} & \geq  0 \: \: \, \qquad \qquad  \forall k=1,...,K \quad\forall i=1,...,n\\
     s_{k,i}+\rho_
   {i}^{(k)}-\rho_
   {i}^{(k-1)}  & \geq  0 \: \: \, \qquad \qquad \forall k=1,...,K \quad \forall i=1,...,n\\															 
              v^{(k)}_j &  \geq  V_{LB}^{(j)}  \: \quad \qquad  \forall k=1,...,K \quad \forall j=1,...,m \\			
			v^{(k)}_j & \leq   V_{UB}^{(j)} \: \quad \qquad  \forall k=1,...,K \quad \forall j=1,...,m\\
             \rho_
   {i}^{(k)} & \in  \{0,1\}  \: \: \: \, \quad \quad  \forall k=1,...,K \quad \forall i=1,...,n \\
		\end{aligned}
	\end{equation} 
  \hrulefill
\vspace*{4pt}
 \end{figure*}

\section{System Model}
\label{sys}















 
The problem introduced in this paper is a thermal-aware workload distribution problem that considers both power consumption and switching costs in the presence of demand variations. We consider a discrete time demand model in which there are $K$ time slots and the demand at time slot $k$ is denoted by $D_k$, the number of required servers at time slot $k$. The problem is a generalized form of the problem introduced in \cite{M3}\cite{our}. For the case of one time slot or fixed demand, presented in \cite{our}, a general nonlinear optimization problem is considered for minimizing the total power consumption in a data center. The system considered in this paper includes $m$ cooling facilities and $n$ servers. The decision variables are the cooling parameters and the server utilizations at time slot $k, k=1,...,K$, denoted by the vectors $v^{(k)}_{m \times 1}$ and $\rho^{(k)}_{n \times 1}$, respectively. As a power reduction scenario, two red-line temperatures are considered corresponding to idle or fully utilized servers, so the server utilizations are 0 or 1. This helps reduce the cooling effort because the lightly loaded servers have a higher red-line temperature. The servers are assumed to be identical. The power consumption and thermal models are generally nonlinear. The cost function is the summation of cooling and IT power consumption along with the cost of workload migration and switching the servers between idle or fully utilized (or on and off states in the case of server consolidation) in consecutive time slots. Addition of the switching cost controls the utilization variation of the servers (and the temperature variations) which as discussed in Section \ref{int}, can lead to increased server lifetimes. So, the trade-off of power consumption and the impact on server reliability due to the frequency of varying server utilizations is considered. There are performance and temperature constraints for each time slot.

We assume that the initial workload distribution is denoted by $\rho^{(0)}$. Thus, the problem that we wish to solve is problem \eqref{p1},
 where $F(v^{(k)})$ is the cooling power consumption corresponding to the cooling variable vector $v^{(k)}_{m \times 1}$ at time slot $k$, $\rho^{(k)}_{n \times 1}$ is the vector of workload distribution at time slot $k$, and $M(v^{(k)},\rho^{(k)})$ is the function corresponding to the thermal model. Within each time slot, a steady-state thermal model is considered. In other words, we assume the time slots are long enough (in the range of minutes) so that a steady-state thermal model is appropriate. The first constraint is a performance constraint with the target demand $D_k$, and the second constraint limits the inlet temperatures to be less than the corresponding red-line temperatures, $T_{idle}$ and $T_{busy}$ (according to \cite{M1}, $T_{idle}>T_{busy}$). The cost of switching (and migration) per server for the $k$th time slot is denoted by $w_k$. The computing (IT) power consumption of server $i$ in the $k$th time slot is denoted by $P(\rho_i^{(k)}, t^{(k)}_i)$, where $t^{(k)}_{n \times 1}=M(v^{(k)},\rho^{(k)})$ is the vector of server inlet temperatures at time slot $k$. The vectors of lower bounds and upper bounds for the cooling variables are
 $V_{LB}$ and $V_{UB}$, respectively. 

 
 There are many possible models that could be used for IT power consumption, but we focus on one choice. We ignore the dependence of IT power consumption on the inlet server temperature. The model is $ P(\rho_i^{(k)})= c+ d \rho_{i}^{(k)}$, where $c$ and $d$ are constants, but we assume that there is server consolidation, so that idle servers are turned off and $P(\rho_i^{(k)})=0$ when $\rho_{i}^{(k)}=0$. In general, server consolidation may change the thermal model but we leave that as a topic for future work.
Server consolidation requires an extra step of linearizing the IT power consumption.

%

  The approach proposed in \cite{our} to solve nonlinear power optimization problems is to linearize the problem and in NP-complete cases (when there are integral variables) propose heuristics developed for approximating the solution of the resulting integer programming problems. The linearized version of the single time slot problem extracted in \cite{our} (with some modifications, assumptions and normalization) is problem \eqref{pr1},
where $a=T_{idle}-T_{busy}>0$, $b=T_{idle}$, $A_{n \times m}$, $B_{n \times n}$ and $E_{n \times 1}$ are the cooling matrix, the heat-recirculation matrix and the constant part, respectively and $A_{i,j}, B_{i,j} \geq 0$ (nonnegative entries). In \cite{our}, we showed that problem \eqref{pr1} is NP-complete and proposed a heuristic to construct an approximate solution.
\begin{figure*}

\begin{equation}\label{p2}
		\begin{aligned}
		 \min  \: \quad \sum_{k=1}^{K}\sum_{j=1}^{m} v^{(k)}_j  + \sum_{k=1}^{K} & w_k \sum_{i=1}^{n} s_{k,i} + 
   \sum_{k=1}^{K} \sum_{i=1}^{n} (\frac{P_{\epsilon}}{\epsilon}\rho_{i}^{-(k)}+  c\rho_{i}^{+(k)}+  d(1-y_{i}^{(k)})) \\
			\text{s.t.} \quad			 
			 \sum_{i=1}^{n} (\rho_
   {i}^{+(k)}+ \rho_{i}^{-(k)})  & \geq  D_k \: \: \qquad \qquad \forall k=1,...,K\\	
						 -\sum_{j=1}^{m}A_{l,j} v_{k,j}+ \sum_{i=1}^{n}B_{l,i} (\rho_
   {i}^{+(k)}+\rho_{i}^{-(k)})	+ a (\rho_
   {l}^{+(k)}+ \rho_{l}^{-(k)}) & \leq b- E_l \quad \qquad \forall k=1,...,K \quad  \forall l=1,...,n \\		
			 s_{k,i}-(\rho_
   {i}^{+(k)}+\rho_{i}^{-(k)})+(\rho_
   {i}^{+(k-1)}+\rho_{i}^{-(k-1)}) & \geq  0  \qquad  \qquad \quad \forall k=1,...,K \quad \forall i=1,...,n\\
     s_{k,i}+(\rho_
   {i}^{+(k)}+\rho_{i}^{-(k)})-(\rho_
   {i}^{+(k-1)} + \rho_{i}^{-(k-1)})  & \geq  0  \qquad \qquad \quad \forall k=1,...,K \quad \forall i=1,...,n\\			
												 \rho_{i}^{+(k)} +  y_i^{(k)} & \leq  1  \qquad \qquad \quad \forall k=1,...,K \quad \forall i=1,...,n \\
              \epsilon y_i^{(k)} +  \rho_{i}^{+(k)}  & \geq  \epsilon \, \qquad \qquad \quad \forall k=1,...,K  \quad \forall i=1,...,n \\
               \epsilon y_i^{(k)} -  \rho_{i}^{-(k)}  & \geq  0 \qquad  \qquad  \quad \forall k=1,...,K \quad \forall i=1,...,n  \\
               v^{(k)}_j & \geq  V_{LB}^{(j)} \: \: \, \qquad \quad \forall k=1,...,K \quad  \forall j=1,...,m \\		
			 v^{(k)}_j  & \leq   V_{UB}^{(j)} \: \: \, \qquad \quad \forall k=1,...,K \quad \forall j=1,...,m \\
              \rho_
   {i}^{+(k)} & \geq  0 \qquad \qquad  \quad \forall k=1,...,K  \quad \forall i=1,...,n \\
               \rho_
   {i}^{-(k)}  & \geq  0 \qquad \qquad \quad \forall k=1,...,K \quad \forall i=1,...,n \\
      y_i^{(k)} & \leq  1 \qquad \qquad  \quad \forall k=1,...,K \quad \forall i=1,...,n  \\
   y_i^{(k)}  & \geq  0   \qquad \qquad  \quad \forall k=1,...,K \quad \forall i=1,...,n \\
				\end{aligned}
	\end{equation} 
   \hrulefill
\vspace*{4pt}
\end{figure*}
Similarly, we first linearize problem \eqref{p1} and then generalize the heuristic 
proposed in \cite{our} to approximate the solution of the linear problem. 
Linearizing the switching cost is straightforward and leads to introducing the new variables $s_{k,i}$. When the server utilizations are 0 or 1, linearizing the IT power consumption is also straightforward. In this case $P(\rho_i^{(k)})= (c+d) \rho^{(k)}_i$. So, the integer linear programming problem is problem \eqref{p3}.


   \begin{figure}
\includegraphics[width=9cm]{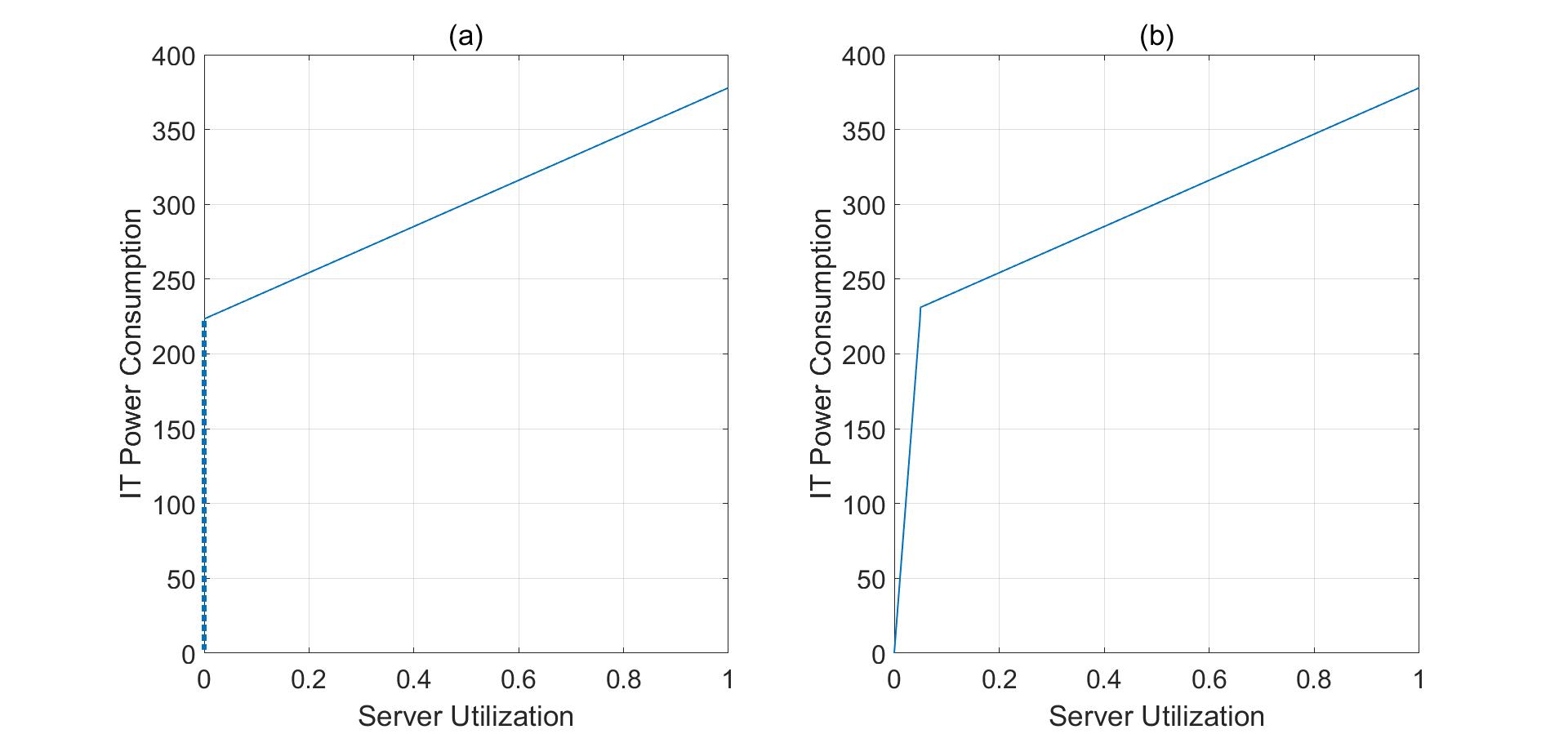}
\centering
\caption{a) The actual IT power consumption (Watts) b) The piecewise linear approximation of IT power consumption}
\label{opt}
\end{figure}

\begin{figure*}[!t]



 \begin{equation}\label{new2}
		\begin{aligned}
		 \min  \quad \sum_{r=1}^{R}  \max \limits_{ l \in S_r} &\frac{[\sum_{i=1}^{n}B'_{l,i} \rho_i^{(k)} -(\sum_{j=1}^{m}A_{l,j} v^{*(k)}_j+b-E_l)]^+}{ z_l} 
 + w_{k} \sum_{i=1}^{n} |\rho_{i}^{(k)}-\hat{\rho_{i}}^{(k-1)}| + w_{k+1} \sum_{i=1}^{n} |\rho_{i}^{*(k+1)}-\rho_{i}^{(k)}| \\
			\text{s.t.} \quad \sum_{i=1}^{n} \rho_i^{(k)} & =  D^*_{k} \\		
			\rho_i^{(k)} & \in   \{0,1\}  \quad  \quad \qquad \forall i=1,...,n \\			 
		\end{aligned}
	\end{equation}
  \hrulefill
\vspace*{4pt}
 \end{figure*}

However, with the relaxation of server utilizations that is needed for the approximation algorithm, more work is needed to linearize the IT power consumption in problem \eqref{p1}. According to the IT power consumption model, in the case of consolidation there is a jump in $P(\rho_i^{(k)})$ when $\rho_i^{(k)}=0$. We approximate $P(\rho_i^{(k)})$ with a piecewise linear function as is shown in Fig.\ \ref{opt}. For a small value $\epsilon$, the IT power consumption is $P_{\epsilon}= c\epsilon+d$. If $\rho_i^{(k)} \leq \epsilon$, then the IT power consumption is approximated as $P(\rho_i^{(k)})= \frac{P_{\epsilon}}{\epsilon} \rho_i^{(k)}$, and if $\rho_i^{(k)} > \epsilon$, then $P(\rho_i^{(k)})=c\rho_i^{(k)}+d$. To linearize these conditions, we divide $\rho_i^{(k)}$ into two parts, $\rho_i^{(k)}=\rho_i^{-(k)}+ \rho_i^{+(k)}$, where one of the following cases is true, depending on the value of the new 0-1 variable $y_{i}^{(k)}$. If $y_{i}^{(k)}=1$, then $\rho_i^{-(k)} \leq \epsilon$ and
    $\rho_i^{+(k)}=0$, otherwise $\rho_i^{-(k)} = 0$ and $ 
    \rho_i^{+(k)} \geq \epsilon$.
       This procedure leads to extra constraints being added to problem \eqref{p3}. Finally the relaxed form of the problem ($0 \leq y_{i}^{(k)} \leq 1$) is problem \eqref{p2}.

 

 \section{Approximation Algorithm}
\label{app}

  Our aim is to approximate the solution of problem \eqref{p3} and use it for the original problem \eqref{p1}. We generalize the H2 heuristic in \cite{our} to approximate the solution of problem \eqref{p3}. The proposed heuristic is based on gradual rounding of the fractional solution of the relaxed linear problem \eqref{p2}. Let us denote the solution of problem \eqref{p2} by $(v^{*(k)}, \rho^{*(k)}), \forall k=1,...,K$. In H2, the main idea for approximating the solution of problem \eqref{pr1} efficiently is to link the original problem to a problem that can be approximated more efficiently. The efficiency comes from reducing the number of constraints and decision variables. The main variables in problem \eqref{pr1} are the server utilizations, because by knowing them, the cooling parameters can be found by solving a standard linear programming problem. In the problem that H2 approximates, the decision variables are only the server utilizations and there is only one constraint, on the number of working servers. In the gradual rounding of the server utilizations, at each step, the decision of which server to be turned off (or keep idle) is made by redistribution of the load of each server to the other servers and calculation of the cost for a new optimization problem that aims at minimizing the total increase (as compared to the fractional cost) in the dominant cooling variables (the dominant cooling variable for server $i$ is the variable with the largest entry in the $i$th row of cooling matrix $A$). That is because the total increase in the dominant cooling variables is assumed to be a good approximation for the total increase in the cooling variables for the original problem \eqref{pr1}. For our problem, the proposed heuristic, called DCVS (Dominant Cooling Variable with Switching cost), is similarly based on gradual rounding of the fractional server utilizations. However, instead of one problem, $K$ problems are approximated. The values of $\hat{\rho}^{(k)}$ are computed consecutively, as the greatest correlation between demands will typically be between consecutive time slots. The problem for time slot $k$ is problem \eqref{new2},
 where $B'=B+I_{n \times n}$ ($I$ is the identity matrix) and there are $R$ dominant cooling variables (the variables with the largest corresponding coefficient for at least one row of $A$). $S_r$ is the set of servers with corresponding dominant cooling variable $r$, $z_l$ is the corresponding coefficient of the cooling variable $r$ (in the $l$th row of $A$) for the server $l \in S_r$ and $D^*_{k} = \lfloor \sum_{i=1}^{n} \rho^{*(k)}_i \rfloor$ ($\lfloor\rfloor$ is the floor function). 
  The cost function for \eqref{new2} is an approximation of the component of the cost function of problem \eqref{p3} that is affected by the value of $\hat{\rho}^{(k)}$. There is no IT power consumption term because it is a constant when the number of working servers is fixed (equal to $D^*_{k}$). The problem for $k=K$ does not include the last term in the cost function.

 Similarly to H2, DCVS is greedy and includes three phases. The first (main) phase is modified to approximate the solution of problem \eqref{p1} in terms of server utilizations, as described in Algorithm \ref{algorithm2}. At each step, the algorithm redistributes the load of each server to other servers (proportional to their current load) and chooses the server to be turned off, based on the cost for problem \eqref{new2}. 
The other two phases can be found in \cite{our}. In phase 2, servers that are turned off in the fractional solution are also considered to be turned on in the integral solution. In phase 3, there are small perturbations of $A$, $B$ and $E$, which lead to multiple fractional solutions where the best solution among them is chosen.

  \section{MPC Approach}
  \label{mpc}

An MPC approach is useful in the presence of demand predictions, although it can also be used when demands are known exactly to reduce the size of the problem.
   We assume that instead of actual demands, we have a noisy version of demands coming from a prediction scheme. In the presence of demand predictions, the weights $w_k$ should be chosen to depend on $k$ as it is reasonable to assume that knowledge about the future demands is more accurate for closer time slots. In general, it is reasonable to assume that $w_k \leq w_{k'}$ when $k \geq k'$.

   In problem \eqref{p3}, it may not be efficient or sufficiently precise to solve the problem for the whole time interval of size $K$. This is both due to the size of the problem and the fact that distant demand predictions may not be sufficiently accurate. One possibility to address these issues is using an MPC approach. The main idea of MPC is considering a window of size $W$ and using the predictions for the next $W$ time slots to compute the workload distribution in the next time slot.
  This reduces the greediness of the algorithm by using the information for several time slots. It also allows for updates to predicted demand values to be considered, each time the solution for the next time slot is calculated. We use the MPC scheme which is described in Algorithm \ref{algorithm3}. Each time, a problem of size $W$ is solved and the solution for the first time slot is kept and used as the initial workload distribution for the next round.

\begin{algorithm}[H]
 \caption{Calculation of $\hat{\rho}^{(k)}, \forall k=1,..., K$}\label{algorithm2}
  \begin{algorithmic}[1] 
  
    \State Solve problem \eqref{p2} and let the solution be $(v^{*(k)},\rho^{*(k)}), \forall k=1,...,K$
    \State $\hat{\rho}^{(0)}=\rho^{(0)}$

    \For {$k=1:K$}
   \State $S= \{i \in \rho^{*(k)}| 0 <\rho^{*(k)}_i \leq 1\}$
   \State $l=|S|-D^*_{k}$
   \State $\hat \rho=\rho^{*(k)}$
   \While {$ l \neq 0$}
   \For {$i \in S$}
   \Comment{distributing the load of server $i$ to the other servers, proportional to their current load}
   \State $\rho^{[i]}=\hat\rho$
   \State $S'=S-\{i\}$
   \State $r= \rho^{[i]}_i $
   \State $\rho^{[i]}_i =0$
   \For {$j \in S'$}
   \State $\rho^{[i]}_j= \rho^{[i]}_j+ \frac{\rho^{[i]}_j}{\sum \limits_{k \in S'}\rho^{[i]}_k} r$
   \EndFor
\While {$\exists s \in S', \rho^{[i]}_s > 1$} 
\Comment{fixing the loads that are greater than 1}
   \State $r=\rho^{[i]}_s -1 $
   \State $\rho^{[i]}_s =1$
\State $S'=S'-\{s\}$
  \For {$j \in S'$}
   \State $\rho^{[i]}_j= \rho^{[i]}_j+ \frac{\rho^{[i]}_j}{\sum \limits_{l \in S'}\rho^{[i]}_l} r$
   \EndFor
\EndWhile
\State $x_i=$ value of the cost function of problem \eqref{new2} for $\rho^{[i]}$
   \EndFor
   \State Remove $i$ with the smallest $x_i$ from $S$ and $\hat \rho=\rho^{[i]}$
\State $l=l-1$
      \EndWhile
            \State $\hat{\rho}^{(k)}=\hat{\rho}$
      \EndFor
\State \textbf{return} $\hat{\rho}^{(k)}, \forall k=1,...,K$
	 \end{algorithmic}
  \end{algorithm}
  
\begin{algorithm}[H]
 \caption{Calculation of $\hat{\rho}^{(s)}, \hat{v}^{(s)}$ using MPC approach with window size $W$}\label{algorithm3}
  \begin{algorithmic}[1] 
\State update the (predicted) demand values
\State solve problem \eqref{p3} for $k=s,..., s+W-1$ and call the solution $(v'^{(k)}, \rho'^{(k)}), \forall k=s,..., s+W-1 $
\State $\hat {\rho}^{(s)}=\rho'^{(s)}, \hat{v}^{(s)} =v'^{(s)}$
\State \textbf{return} $\hat {\rho}^{(s)}$ and $\hat{v}^{(s)}$
	 \end{algorithmic}
  \end{algorithm}

 \section{Evaluation}
 \label{eva}


\begin{table*}
  \begin{center}
    \caption{Performance of the Algorithms for Different Values of $w$}
    \label{t1}
    \begin{tabular}{|ccccccccccccc|} 
\hline


&& \multicolumn{2}{c}{OPTi}&&\multicolumn{2}{c}{Sep}&&\multicolumn{2}{c}{SR}&&\multicolumn{2}{c|}{DCVS} \\

\hline

$w$&&avg&wrc&&avg&wrc&&avg&wrc&&avg&wrc\\
\hline

1 && 
1.01 &  1.02  &&  
1.01  &  1.10   &&  
1.02  &  1.05  &&  
1.01  &  1.05       \\

2 && 
1.01 &  1.02  &&
1.01 &  1.06   &&    
1.02 &  1.06  &&  
1.01 &  1.06     \\

4 && 
1.01 & 1.02 && 
1.01 & 1.10   &&  
1.02 & 1.04   &&  
1.01 & 1.07      \\

8 && 
1.01 &  1.02  && 
1.02 &  1.06  &&  
1.02 &  1.05  &&  
1.01 &  1.07    \\

16 && 
1.01   &  1.02   &&  
1.01 &  1.08    && 
1.02 &  1.05    &&  
1.01 &  1.06  \\
 



32 && 
1.01 &  1.02   &&  
1.01  &  1.07   &&  
1.01  &  1.05     &&  
1.01   &  1.08      \\

64 && 
1.01 &  1.02   &&  
1.02 &  1.13   &&  
1.01 &  1.04   &&  
1.01 &  1.05   \\

128 && 
1.01  &  1.01   &&  
1.02  &  1.13   &&  
1.01 &  1.05   &&  
1.01    &  1.05     \\

512 && 
1.01 &  1.01    &&  
1.04 &  1.24    &&  
1.01 &  1.04    &&  
1.01 &  1.04  \\

1024 && 
1.00 &  1.01  &&  
1.11 &  1.55     &&  
1.01 &  1.15     &&  
1.00 &  1.04  \\



 
2048 && 
1.00 & 1.01   && 
1.21 & 1.87   &&  
1.01    & 1.16    && 
1.01   & 1.02        \\
 
4096 && 
1.00 &  1.01  &&   
1.36 &  2.04  &&  
1.01 &  1.09  &&   
1.00 &  1.03   \\
 
8192 &&
1.00 &  1.00  &&  
1.49 &  2.63  &&  
1.01 &  1.16  &&  
1.00 &  1.02   \\
 
16384 && 
1.00 &  1.00  && 
1.56 &  2.81  && 
1.00 &  1.18  && 
1.00 &  1.01    \\
 
32768 && 
1.00 &  1.00  &&  
1.63 &  2.96  &&  
1.01 &  1.23  && 
1.00 &  1.00   \\
 

   \hline

    \end{tabular}
  \end{center}
\end{table*}

\begin{figure}
\includegraphics[width=9cm]{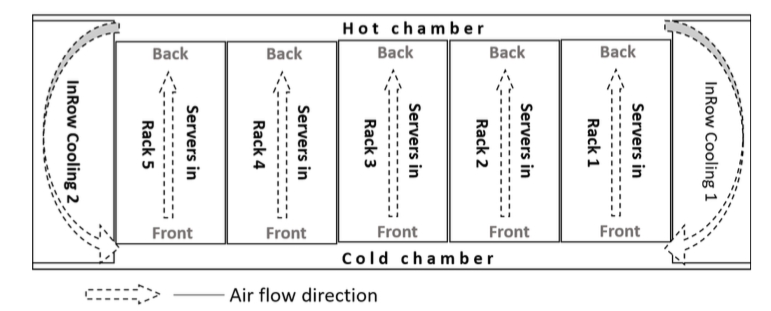}
\centering
\caption{The data center's top view according to \cite{roh}}
\label{opt2}
\end{figure}

 The system we use for evaluation comes from an experimental data center at McMaster University that is modeled in \cite{roh}. The data center has 25 servers located in 5 racks and two cooling facilities. The cooling variables are the chilled water temperature and total air flow generated by two fans at either end of the racks. The top view of the data center is shown in Fig.\ \ref{opt2}. The functions $M$ and $F$ in problem \eqref{p1} are simulated based on the model in \cite{roh} for inlet temperatures and cooling power consumption. The platform we used was MATLAB R2021b running on a 64-bit system with an i7-1185G7 processor and 8-GB RAM. The function $M$ is not explicitly given and the inlet temperatures are calculated based on solving a set of differential equations, an operation that is computationally intensive (each call takes around 1.4 seconds). The next step is regression on the functions $F$ and $M$ to linearize the problem. The function \textit{regress} in MATLAB is used and the data points are uniformly at random chosen from the defined ranges for cooling variables and server utilizations (the server utilizations are continuous). We set $T_{idle}=35$ and $T_{busy}=27$ (degrees Celsius), $V_{LB}=[1300, 10]$ and $V_{UB}=[2300, 20]$. Additional details are provided in \cite{our}. So, the matrices $A$, $B$ and $E$ in problem \eqref{p3} are known. We perform a (small) random perturbation of the matrices, each time that the algorithms are run. We assume the IT power consumption model is according to server consolidation with coefficients also coming from the  model in \cite{roh}.
  In the simulation results presented in \cite{our}, we have shown that the solution of the linear system approximated by the proposed heuristic works well for the original nonlinear system. So, here we focus on the evaluation of the heuristics for the linearized system. We also assume $w_k=w, \forall k=1,...,K$.
 
 We use simple rounding (SR) as the baseline algorithm. In simple rounding for each time slot $k$, the $D_k^*$ largest values in $\rho^{*(k)}$ are rounded to one. We also solve the single time slot problem for each of the $K$ time slots (without switching cost) using the \textit{intlinprog} function in MATLAB and calculate the cost of the solution for the multiple time slot problem (with switching cost). This scheme is called Sep in the results. For this data center example, we show that although the performance of SR and H2 are very close for the one time slot problem according to \cite{our}, for the multiple time slot problem DCVS clearly works better. 
 
 The evaluation includes three parts. In the first part, the sensitivity of the approaches to the value of $w$ is evaluated.  In the second part, the performance is evaluated in the presence of demand fluctuations with different patterns. Finally, the performance of the integrated MPC approach is evaluated for the actual and noisy demand values. We present the average and the worst case ratios (avg and wrc columns in the results) when the solution is compared with the solution of the relaxed problem \eqref{p2}.

 
The first results correspond to sensitivity to $w$. As $w$ increases, the switching cost becomes more dominant. Starting from $w=1$, the value of $w$ doubles. The number of intervals $K$ is equal to 3. The pair of demands $(D_1,D_3)$ covers all possible combinations, where the values for the demand are chosen from $D=\{1,2, .... , 24\}$. For each combination, $D_2$ is randomly chosen from $D$. The results are reported in Table \ref{t1}, with an extra column OPTi corresponding to solving the problem using the \textit{intlinprog} function in MATLAB. Although OPTi has the best performance, in \cite{our} we showed that the running time does not scale well for larger problem sizes. The results show that the performance of DCVS is more resilient to changes in $w$ and for larger values of $w$, SR has poor performance with respect to the worst case ratio. The reason is that for the fractional optimal solution, the largest $D^*_k$ utilizations in time slot $k$ are not necessarily a subset of the largest $D^*_{k+1}$ in time slot $k+1$ or vice versa. So, when the switching cost is very large, SR leads to large costs due to the switching cost component. For example, we saw this effect for the input $[7, 9, 9]$ and $w=1000$, when we used SR for several systems that are perturbed versions of the main system introduced in this section. This effect was seen for roughly one out of 10 of these systems. The utilizations for one of the servers were the same for the first and second time slots but SR rounded the first to 1 and the second to 0.

Another observation from Table \ref{t1} is that although the average ratio is better for DCVS, for smaller values of $w$, the worst case ratio is more for DCVS as compared to SR. When DCVS redistributes the load, it considers both the cooling power consumption and the switching cost, so in the process of rounding, some redistributions that generate lower switching cost may be chosen although they have greater cooling power consumption. This may be problematic because when the rounding is complete, the switching cost may be the same for other redistributions with lower cooling power consumption, although the switching cost was greater in the process. We saw this effect for the demand sequence $[1,3,1]$ and $w=4$.
So, the greediness of DCVS may be problematic in some cases. The results also show the performance of the Sep scheme is not as good as the others, specially for larger values of $w$, as a result of Sep ignoring correlations between consecutive time slots providing opportunities to reduce switching costs.

 In the second part, we evaluate the performance of the algorithms in the presence of workload fluctuations with different rates. The number of time slots is $K=9$. For each round of simulations, three values are chosen from $D$. So, there are three possibilities for the demand values. The rate of fluctuations is then controlled by a probability $p \in \{0.1, 0.5 , 0.9\}$. $D_1$ is chosen randomly from the set of three demand values. For the next demand values $D_2$ to $D_9$, we pick the previous demand value with probability $1-p$ or pick a different value randomly with probability $p$. This procedure is repeated 100 times for each value of $p$ in each round of simulations. We also set $w=1000$. There are 20 round of simulations as reported in Table \ref{t2}. According to the results in Table \ref{t2}, DCVS has the best performance and is more resilient to the workload fluctuations with different rates. In addition, Table \ref{t3} reports the running times for $K=20$, where at each round of simulations (each row in Table \ref{t3}) the demand sequences are generated as explained for the previous results in Table \ref{t2}. The results show that the running times for DCVS and SR are close and both are clearly faster than the \textit{intlinprog} function in MATLAB. 

\begin{table}
  \begin{center}
    \caption{Performance of the Algorithms in the Presence of Workload Fluctuations with Different Rates}
    \label{t2}
    \begin{tabular}{|cccccccc|} 
\hline


 \multicolumn{2}{|c}{Sep}&&\multicolumn{2}{c}{SR}&&\multicolumn{2}{c|}{DCVS} \\

\hline

avg&wcr&&avg&wcr&&avg&wrc\\
\hline

1.49  &  2.26   &&  
1.01  &  1.06   &&  
1.01  &  1.02      \\

1.50 &  2.43   &&    
1.01 &  1.05  &&  
1.01 &  1.01     \\

1.71 & 2.51   &&  
1.00 & 1.07   &&  
1.00 & 1.02     \\

1.46 &  2.33  &&  
1.01 &  1.03   &&  
1.01 &  1.01    \\

1.45 &  1.99     && 
1.01 &  1.03     &&  
1.01 &  1.02   \\
 



1.61  & 2.25    &&  
1.01   &  1.04      &&  
1.01  &  1.02      \\

1.42 &  2.23   &&  
1.01 &  1.03   &&  
1.01 &  1.02    \\

1.34 &  1.84   &&  
1.01 &  1.03    &&  
1.01    &  1.01     \\

1.51 &  2.09    &&  
1.01 &  1.06   &&  
1.01 &  1.02  \\

1.60 &  2.24     &&  
1.01 &  1.04     &&  
1.01 &  1.02  \\




1.40 & 2.03   &&  
1.01    & 1.05     &&  
1.01    & 1.02       \\

1.64 &  2.13  &&  
1.01 &  1.05  &&   
1.01 &  1.01    \\

1.48 &  2.26  &&  
1.01 &  1.08  &&  
1.01 &  1.02  \\

1.44 &  1.92  && 
1.01 &  1.05  && 
1.01 &  1.02    \\

1.47 &  2.22  &&  
1.01 &  1.04  && 
1.01 &  1.01   \\

1.32 &  1.71  &&  
1.01 &  1.03  && 
1.01 &  1.01   \\

1.44 &  2.01  &&  
1.01 &  1.05  && 
1.01 &  1.01   \\

1.48 &  2.48  &&  
1.01 &  1.03  && 
1.01 &  1.03   \\

1.35 &  1.75  &&  
1.01 &  1.03  && 
1.01 &  1.02   \\

1.58 &  2.31  &&  
1.01 &  1.03  && 
1.01 &  1.03   \\
 

   \hline

    \end{tabular}
  \end{center}
\end{table}

The final results correspond to the integrated MPC approach. 
 The number of time slots is $K=50$ and the size of the planning window $W$ is varied between 1 and 10. To calculate the solution over $K=50$ time slots, the MPC approach uses a total of $K+W-1$ demand values. So with $W_{max}=10$, the length of the required demand sequence is $50+10-1=59$. We consider six scenarios for generation of demand sequences.
 There are three cases for the range of demand values. Case 1 corresponds to choosing the demand values from $D=\{1,..., 24\}$ uniformly at random. In Case 2, the next demand $D_{k+1}$ is chosen from
 the range $[max(D_k-5, 1), min(D_k+5, 24)]$. In Case 3, the range for choosing $D_{k+1}$ is $[max(D_k-2, 1), min(D_k+2, 24)]$. There is also a probability $p$ that is the probability of changing the demand value for the next time slot. We chose two values of 0.2 and 0.8 for $p$. For example, in Case 3 and $p=0.2$, with probability 0.2, the next demand value $D_{k+1}$ is different from $D_k$ and it is chosen from the range $[max(D_k-2, 1), min(D_k+2, 24)] / \{D_k\}$. 
We also set $w=1000$. The total optimal cooling power consumption for each time slot ranges between 1500 and 2000. The IT power consumption for each working server is also around 375 Watts (223.4 + 154.5).  The simulation is repeated 10 times for each scenario. For each round, the values in $\rho^{(0)}$ are chosen randomly from $\{0,1\}$.  

\begin{table}
  \begin{center}
    \caption{Running Time of the Algorithms (in seconds) in the Presence of Workload Fluctuations with Different Rates}
    \label{t3}
    \begin{tabular}{|ccccc|} 
\hline


 \multicolumn{1}{|c}{OPTi}&&\multicolumn{1}{c}{SR}&&\multicolumn{1}{c|}{DCVS} \\

\hline


17.53  &&  
0.56  &&  
0.71\\

3.08  &&  
0.68  &&  
0.78\\

8.7  &&  
0.68  &&  
0.87\\

10.55  &&  
0.61  &&  
0.79\\

6.58  &&  
0.70  &&  
0.87\\



10.14  &&  
0.65  &&  
0.81\\

15.42  &&  
0.73  &&  
0.98\\

10.68  &&  
0.61  &&  
0.80\\

8.83  &&  
0.60&&  
0.77\\

3.10  &&  
0.61  &&  
0.80\\
 


 



   \hline

    \end{tabular}
  \end{center}
\end{table}

In addition to using the actual demands, we assume that we have a noisy version of demands coming from demand predictions. For a window of size $W$, starting from the time slot $s$,  
we assume that $D_s$ is the actual value. However for $D_{s+1}, ..., D_{s+W-1}$, noise is added to the actual demand. The value of noise for the time slot $s+k-1, k=2,...W,$ is randomly chosen from the interval $[-\eta \times k, \eta \times k]$, where $\eta$ is the basic noise value. So, when the window shifts to the next time slot, the added noise is resampled with an updated distribution because $k$ changes to $k-1$ for the same time slot ($s$ changes to $s+1$). We apply the floor function to the noise value and add it to the actual demand. If the predicted demand is less than 1 or greater than 24, we choose the values 1 or 24, respectively. We consider three cases of $\eta=0, \eta=1, \eta=3$, where $\eta=0$ corresponds to the actual values without noise. For each value of $\eta>0$ we repeat the procedure five times. The values reported in Table \ref{t4} are the ratios to the cost calculated for the whole interval with actual demand values by using DCVS. The reference cost is very close to  optimal.

When $w$ is very small, the correlation between time slots is small, so using the MPC approach is not necessary. In this case, the problem can be solved by examining a single time slot problem. When $w$ is larger, it becomes more important that the working servers in consecutive time slots are correlated. First, the load for each time slot is equal to the corresponding demand and in consecutive time slots, the set of servers with the smaller demand is a subset of the set of servers with larger demand. However, for larger values of $w$, to decrease the switching cost, the number of working servers may also be greater than the corresponding demand. Finally, as $w$ becomes very large, the working servers for all time slots are the same. In this case, it is only necessary to focus on the time slot with the largest demand and use that solution for all other time slots. We expect that using the MPC approach is beneficial when there is an actual trade-off between the switching costs and the other components of the cost function.  

 The results for $w=1000$ are shown in Table \ref{t4}. We focused on $w=1000$ because it clearly shows the trade-off between the costs and the performance of the MPC approach. The results for $\eta=0$, show that for smaller window sizes (in particular $W=1$), the performance is poor for all scenarios, with good performance achieved when $W=4$. We can infer that in the short term the switching cost may be dominant. 
 However, when the size of $W$ increases the IT (and cooling) power consumption does not allow extra servers to be working in several time slots. As an example, for demand $[15, 5, 5]$, to decrease the switching cost, it may be beneficial to increase the number of working servers at the second and third time slots above 5, however for the demand $[15, 5, 5, 5, 5, 5, 15]$, it might not be beneficial to increase the load in all time slots with demand equal to 5, because of the increase in IT (and cooling) power consumption. 
 In general, the long term and short term solutions may be different. It can be inferred that as long as the window size is not too short, the MPC approach is beneficial as is shown for the case of $W=3$ or $W=4$. Using a window size of 4 (or 3) is also acceptable in the presence of noise. Using larger window sizes may not be helpful, in particular in the presence of noise when it might even degrade the performance, since the predicted demand is far from the actual demand. We also see how the performance degrades when the noise increases. So, updating information in the MPC approach is also helpful to improve the performance.
\begin{table}[H]
  \begin{center}
    \caption{Performance of the Integrated MPC Approach with DCVS}
    \label{t4}
    \begin{tabular}{|cccccccccc|} 
\hline


\multicolumn{10}{|c|}{Case 1 with $p=0.2$} \\
\hline 
&& \multicolumn{2}{c}{$\eta$=0}&&\multicolumn{2}{c}{$\eta$=1}&&\multicolumn{2}{c|}{$\eta$=3} \\

\hline

$W$&&avg&wrc&&avg&wrc&&avg&wrc\\
\hline

1 && 

1.27  &  1.60   &&  
1.27  &  1.60  &&  
1.27  &  1.60       \\

2 && 
1.09 &  1.18   &&    
1.09 &  1.18  &&  
1.09 &  1.18     \\

3 && 
1.00 & 1.02   &&  
1.01 & 1.02   &&  
1.02 & 1.07      \\

4 && 
1.00 &  1.01  &&  
1.01 &  1.02  &&  
1.03 &  1.07    \\

5 && 
1.00 &  1.01    && 
1.01 &  1.03    &&  
1.04 &  1.11  \\
 





8 && 
1.00  &  1.01   &&  
1.01 &  1.03   &&  
1.04    &  1.12     \\


10 && 
1.00 &  1.01    &&  
1.01 &  1.03     &&  
1.04 &  1.11  \\

\hline

\multicolumn{10}{|c|}{Case 1 with $p=0.8$} \\
\hline 



1 && 

1.10  &  1.19   &&  
1.10  &  1.19  &&  
1.10  &  1.19       \\

2 && 
1.05 &  1.06   &&    
1.05 &  1.07  &&  
1.05 &  1.07     \\

3 && 
1.04 & 1.07   &&  
1.05 & 1.09   &&  
1.08 & 1.14      \\

4 && 
1.02 &  1.04  &&  
1.04 &  1.06  &&  
1.06 &  1.09    \\

5 && 
1.02 &  1.03    && 
1.03 &  1.06    &&  
1.05 &  1.08  \\
 





8 && 
1.02  &  1.02   &&  
1.03 &  1.04  &&  
1.05    &  1.09     \\


10 && 
1.02 &  1.03    &&  
1.03 &  1.04     &&  
1.05 &  1.08  \\

\hline

\multicolumn{10}{|c|}{Case 2 with $p=0.2$} \\
\hline 



1 && 

1.51  &  2.45   &&  
1.51  &  2.45  &&  
1.51  &  2.45       \\

2 && 
1.09 &  1.46   &&    
1.09 &  1.46  &&  
1.09 &  1.47     \\

3 && 
1.01 & 1.02   &&  
1.01 & 1.02   &&  
1.02 & 1.05      \\

4 && 
1.00 &  1.01  &&  
1.01 &  1.02  &&  
1.02 &  1.11    \\

5 && 
1.00 &  1.01    && 
1.01 &  1.03    &&  
1.03 &  1.12  \\
 





8 && 
1.00  &  1.01  &&  
1.01 &  1.06   &&  
1.04    &  1.21     \\


10 && 
1.00 &  1.01    &&  
1.01 &  1.04     &&  
1.04 &  1.26  \\

\hline

\multicolumn{10}{|c|}{Case 2 with $p=0.8$} \\
\hline 



1 && 

1.25  &  1.69   &&  
1.25  &  1.69  &&  
1.25  &  1.69       \\

2 && 
1.06 &  1.15  &&    
1.06 &  1.15  &&  
1.07 &  1.15     \\

3 && 
1.02 & 1.04   &&  
1.03 & 1.06   &&  
1.05 & 1.09      \\

4 && 
1.02 &  1.03  &&  
1.03 &  1.04  &&  
1.04 &  1.07    \\

5 && 
1.02 &  1.04    && 
1.03 &  1.05    &&  
1.04 &  1.08  \\
 





8 && 
1.02  &  1.03   &&  
1.03 &  1.05   &&  
1.04    &  1.09     \\


10 && 
1.02 &  1.04    &&  
1.03 &  1.05     &&  
1.04 &  1.07  \\

\hline

\multicolumn{10}{|c|}{Case 3 with $p=0.2$} \\
\hline 



1 && 

1.41  &  1.98   &&  
1.41  &  1.98  &&  
1.41  &  1.98       \\

2 && 
1.04 &  1.14   &&    
1.04 &  1.18  &&  
1.05 &  1.25     \\

3 && 
1.00 & 1.01   &&  
1.01 & 1.02   &&  
1.01 & 1.04      \\

4 && 
1.00 &  1.01  &&  
1.01 &  1.03  &&  
1.02 &  1.07    \\

5 && 
1.00 &  1.01    && 
1.01 &  1.02    &&  
1.02 &  1.08  \\
 





8 && 
1.00  &  1.01   &&  
1.01 &  1.04   &&  
1.02    &  1.09     \\


10 && 
1.00 &  1.01    &&  
1.01 &  1.02     &&  
1.03 &  1.12  \\

\hline

\multicolumn{10}{|c|}{Case 3 with $p=0.8$} \\
\hline 



1 && 

1.31  &  1.52   &&  
1.31  &  1.52  &&  
1.31  &  1.52       \\

2 && 
1.04 &  1.08   &&    
1.05 &  1.10  &&  
1.05 &  1.10     \\

3 && 
1.00 & 1.02   &&  
1.03 & 1.06   &&  
1.04 & 1.08      \\

4 && 
1.01 &  1.03  &&  
1.03 &  1.06  &&  
1.04 &  1.08    \\

5 && 
1.01 &  1.03    && 
1.02 &  1.05    &&  
1.04 &  1.09  \\
 





8 && 
1.01  &  1.03   &&  
1.02 &  1.05   &&  
1.03    &  1.07    \\


10 && 
1.01 &  1.03    &&  
1.03 &  1.05     &&  
1.04 &  1.06  \\

\hline








 













    \end{tabular}
  \end{center}
\end{table}


 \section{Conclusion}
 \label{con}

In this work, we formulated a nonlinear optimization problem for data centers that considers the switching costs in addition to cooling and IT power consumption. Workload transitions among the servers due to dynamic demands are not beneficial in terms of the switching costs, however they may decrease the power consumption. Next, we used a linearization approach proposed in \cite{our} to approximate the solution. The steps were linearization of the problem and the development of a heuristic to approximate the solution of the linear problem. Finally, we proposed an integrated MPC approach with our proposed heuristics which is helpful to decrease the size of the problem and to incorporate demand predictions. The simulation results show that the proposed schemes are helpful to find a near-optimal solution efficiently. We showed that using an appropriate window size in the MPC approach is important and beneficial. As future work, integrating transient thermal models with dynamic demands would be of interest. Modifying the proposed heuristic to address more problems, for example the case of heterogeneous data centers, is also a possibility. A procedure should also be defined for determining appropriate weights for switching costs. Finding a suitable window size for the MPC approach also needs more investigation.

\end{document}